\documentclass[preprint,eqsecnum,prb]{revtex4}
\usepackage{epsfig,amsmath,bbm,graphicx}
\newcommand{\ee}{\end{equation}}
\newcommand{\be}{\begin{equation}} 
\newcommand{\bea}{\begin{eqnarray}}
\newcommand{\eea}{\end{eqnarray}}
\newcommand{\bml}{\begin{subequations}} 
\newcommand{\eml}{\end{subequations}}

\begin{document}

\title{A multilayer multiconfigurational time-dependent Hartree study of vibrationally coupled electron transport 
using the scattering state representation}

\author{Haobin Wang}
\affiliation{Department of Chemistry and Biochemistry, MSC 3C, New Mexico State
University, Las Cruces, NM 88003, USA}
\affiliation{Beijing Computational Science Research Center,
No. 3 He-Qing Road, Hai-Dian District, Beijing 100084, P.R. China}

\author{Michael Thoss}
\affiliation{Institute for Theoretical Physics and Interdisciplinary Center for Molecular Materials,
  Friedrich-Alexander-Universit\"at Erlangen-N\"urnberg,
  Staudtstr.\ 7/B2, D-91058, Germany}


\begin{abstract}
\baselineskip6mm

The multilayer multiconfiguration time-dependent Hartree method is employed to
study vibrationally coupled charge transport in models of single molecule junctions. To increase the
efficiency of the simulation method, a representation of the Hamiltonian in
terms of the scattering states of the underlying electronic Hamiltonian is
used. It is found that with an appropriate choice of the scattering states 
the artificial electron correlation present in the original representation of
the model is greatly reduced.  This allows efficient simulation 
of the steady-state currents in a wide physical parameter
space, which is demonstrated by several
numerical examples.

\end{abstract}
\maketitle

\section{Introduction}

Charge transport in single-molecule junctions has raised a great deal of interest
recently.\cite{ree97:252,joa00:541,Nitzan01,nit03:1384,Cuniberti05,Selzer06,Venkataraman06,Chen07,Galperin08b,Cuevas10}  Experimental studies on transport properties of molecular junctions revealed
a variety of interesting phenomena, e.g., Coulomb blockade,\cite{par02:722} 
Kondo effect,\cite{lia02:725} negative differential resistance,\cite{che99:1550,Gaudioso00,Osorio10} 
switching and hysteresis.\cite{blu05:167,Riel06,Choi06}
This has stimulated many theoretical developments for understanding quantum
transport at the molecular scale.  Examples of approximate methods employed in this regard  
include the scattering  
theory,\cite{Bonca95,Ness01,Cizek04,Cizek05,Toroker07,Benesch08,Zimbovskaya09,Seidemann10} 
nonequilibrium Green's function (NEGF) 
approaches,\cite{Flensberg03,Mitra04,Galperin06,Ryndyk06,Frederiksen07,Tahir08,Haertle08,Stafford09,Haertle09} 
and master equation methods.\cite{May02,Mitra04,Lehmann04,Pedersen05,Harbola06,Zazunov06,Siddiqui07,Timm08,May08,May08b,Leijnse09,Esposito09,Volkovich11,Haertle11}
To the other end, a variety of (in principle) numerically exact methods have been developed 
to obtain more reliable results for nonequilibrium transport in model systems. These include
the numerical path  integral approach,\cite{muh08:176403,wei08:195316,Segal10} real-time 
quantum Monte Carlo simulations,\cite{Werner09,Schiro09} the numerical renormalization 
group approach,\cite{and08:066804}, the time-dependent density matrix renormalization 
group approach.\cite{HeidrichMeisner09}, and the hierarchical equations of motion
method.\cite{Zheng09,Jiang12}  Our work in this direction is the development of the multilayer 
multiconfiguration time-dependent Hartree (ML-MCTDH) theory in the second quantization representation 
(SQR),\cite{wan09:024114} a systematic, numerically exact methodology to study quantum dynamics 
and quantum transport including many-body effects. For a generic model of vibrationally coupled
electron transport, we have demonstrated\cite{Wang11} the importance of treating the vibronic 
coupling accurately.  Comparison with approximate methods such as NEGF revealed the necessity 
of employing accurate methods such as the ML-MCTDH-SQR approach, in particular in the strong coupling regime. 

The ML-MCTDH-SQR method\cite{wan09:024114} employs a recursive, layered representation of the overall 
wave function and captures correlation effects through a converged multiconfigurational expansion 
in all the corresponding layers.  Like any other methods, the efficiency of the ML-MCTDH-SQR theory 
depends on how the problem is approached, e.g., the choice of the an appropriate coordinate system.  
Consequently, the term ``correlation'' also depends on such a choice.  For example, the quantum
dynamics of a quadratic Hamiltonian, where a full Hessian is present, is highly correlated.  On the
other hand, when this Hamiltonian is transformed to the normal mode representation, there is no
correlation at all between different degrees of freedom.  In this sense we can term the former 
correlation as  ``artificial correlation'' due to the inappropriate representation of the 
Hamiltonian. Finding an optimal representation is thus crucial for solving the problem efficiently.  
Unfortunately, this is typically not an automatic procedure and requires a
clear physical understanding  of the 
problem.  In most cases, a simple analytic reduction of the problem as the
example above is not available.  Instead, 
one has to transform the Hamiltonian according to the physical intuition and known experiences.  One 
of the most frequently used criterion is to ensure that this transformation reduces the problem in 
a certain physical limit.

In a previous ML-MCTDH-SQR study of vibrationally coupled electron transport,\cite{wan09:024114} we had found that 
in some regimes the configuration space required to achieve convergence is quite large.  We noticed,
however, that these regimes are not necessarily  characterized by a strong correlation in the usual sense.  
In fact, it was often relatively easy to converge the simulation when the electron-nuclear coupling is 
strong (i.e., easier to capture the ``real'' correlation effect induced by
electronic-vibrational coupling), but not otherwise 
(i.e. more difficult to capture artificial electron correlation effects), 
especially when the source-drain bias voltage is high.  This has motivated us to examine other 
representations of the Hamiltonian for studying nonequilibrium  charge
transport.  One promising choice is to employ the scattering states of the
underlying electronic  Hamiltonian. The resulting scattering states
representation of the model has been used before to describe quantum transport
in combination with different methods.\cite{Hershfield93,Doyon06,Han07,Oguri07,Anders08,Gelin09} 
In this paper we discuss our implementation of the scattering state representation within the
framework of the ML-MCTDH-SQR theory.  

The paper is organized as follows. Section~\ref{modeltight} outlines the original, tight-binding 
type model for describing the vibrationally coupled electron transport.  
Section~\ref{modelscatter} then discusses the scattering state representation of the same problem
and our specific choice of the reference frame.  Section~\ref{results} illustrates the
performance of our approach by numerical examples for a variety of parameter regimes.
Section~\ref{conclusions} concludes with a summary.

\section{Model for Vibrationally Coupled Electron Transport}\label{modeltight}

\subsection{Model Hamiltonian}

A model that is frequently adopted to study vibrationally coupled electron transport through a 
single-molecule junction is based on tight-binding description for the two metal leads.
It comprises one discrete electronic state at the molecular bridge, two electronic continua 
describing the left and the right metal leads, respectively, and a distribution of harmonic 
oscillators that models the vibrational modes of the molecular bridge.  
The Hamiltonian reads
\begin{subequations}\label{Htot}
\begin{equation}
	\hat H = \hat H_{\rm el} + \hat H_{\rm el-nuc},
\end{equation}
where $\hat H_{\rm el}$ and $\hat H_{\rm el-nuc}$ describe the pure electronic degrees of freedom 
and the nuclear vibrations with their coupling terms to the electronic part, respectively
\begin{eqnarray}\label{Hel}
	\hat H_{\rm el} &=& E_d d^+ d + \sum_{k_L} E_{k_L} c_{k_L}^+ c_{k_L}
	+ \sum_{k_R} E_{k_R} c_{k_R}^+ c_{k_R} \\ \nonumber
  &&	+ \sum_{k_L} V_{dk_L} ( d^+ c_{k_L} + c_{k_L}^+ d )
	+ \sum_{k_R} V_{dk_R} ( d^+ c_{k_R} + c_{k_R}^+ d ),
\end{eqnarray}
\begin{equation}\label{Hel-nuc}
	\hat H_{\rm el-nuc} = \frac{1}{2} \sum_j ( P_j^2 + \omega_j^2 Q_j^2 ) + d^+ d \sum_j 2 c_j Q_j.
\end{equation}
\end{subequations}
Thereby,
$d^+/ d$, $c_{k_L}^+/ c_{k_L}$, $c_{k_R}^+/ c_{k_R}$ are the fermionic creation/annihilation operators 
for the electronic states on the molecular bridge, the left and the right leads, respectively.  
The corresponding electronic energies $E_{k_L}$, $E_{k_R}$ and the molecule-lead coupling strengths 
$V_{dk_L}$, $V_{dk_R}$, are defined through the energy-dependent level width functions
\begin{equation}
	\Gamma_L (E) = 2\pi \sum_{k_L} |V_{dk_L}|^2 \delta(E-E_{k_L}), \hspace{1cm}
	\Gamma_R (E) = 2\pi \sum_{k_R} |V_{dk_R}|^2 \delta(E-E_{k_R}).
\end{equation}
In principle, the parameters of the model can be obtained for a specific
molecular junction employing first-principles electronic structure
calculations.\cite{Benesch09} In this paper, which focuses on the methodology
and general transport properties, however, we will use a generic parameterization.
Employing a tight-binding model, the function $\Gamma (E)$ is given as
\begin{subequations}\label{tight}
\begin{equation}
        \Gamma (E) = \left\{ \begin{array}{ll} \frac{\alpha_e^2}{\beta_e^2} \sqrt{4\beta_e^2-E^2} 
		\hspace{1cm} & |E| \leq 2 |\beta_e| \\
                0 \hspace{1cm} &  |E| > 2 |\beta_e| \end{array}  \right.,
\end{equation}
\begin{equation}
	\Gamma_L (E) = \Gamma (E-\mu_L), \hspace{1cm}  \Gamma_R (E) = \Gamma (E-\mu_R),
\end{equation}
\end{subequations}
where $\beta_e$ and $\alpha_e$ are nearest-neighbor couplings between two lead sites 
and between the lead and the bridge state, respectively.  I.e., the width functions for 
the left and the right leads are obtained by shifting $\Gamma(E)$ relative to the chemical potentials 
of the corresponding leads.  We consider a simple model of two identical leads, in which the chemical 
potentials are given by 
\begin{equation}
	\mu_{L/R} = E_f \pm V/2,
\end{equation}
where $V$ is the source-drain bias voltage and $E_f$ the Fermi energy of the leads. Since only the 
difference $E_d - E_f$ is physically relevant, we set $E_f = 0$.

Similarly, the frequencies $\omega_j$ and electronic-nuclear coupling
constants $c_j$ of the vibrational modes of the molecular junctions are modeled by a spectral density 
function\cite{leg87:1,Weiss93}
\begin{equation}
\label{discrete}
        J(\omega) = \frac{\pi} {2} \sum_{j} \frac{c_{j}^{2}} {\omega_j}
        \delta(\omega - \omega_{j}).
\end{equation}
In this paper, the spectral density is chosen in Ohmic form with an exponential cutoff
\begin{equation}
\label{ohmic}
        J_{\rm O}(\omega)  = \frac{\pi}{2} \alpha \omega e^{-\omega/\omega_c},
\end{equation}
where $\alpha$ is the dimensionless Kondo parameter.  

Both the electronic and the vibrational continua can be discretized by choosing a 
density of states $\rho_e(E)$ and a density of frequencies 
$\rho(\omega)$ such that\cite{tho01:2991,wan01:2979,wan03:1289}
\begin{subequations}
\begin{equation}
	\int_0^{E_k} dE \; \rho_e(E) = k, \hspace{.5in} 
	|V_{dk}|^2 =  \frac{\Gamma(E_k)}{2\pi \rho_e(E_k)}, \hspace{.5in}
	 k = 1,...,N_e,
\end{equation}
\begin{equation}
	\int_0^{\omega_j} d\omega \; \rho(\omega) = j, \hspace{.5in}  
	\frac{c_{j}^{2}} {\omega_j} = \frac{2}{\pi} \frac{J_{\rm O}(\omega_j)}{\rho(\omega_j)},
 	\hspace{.5in} j = 1,...,N_b.
\end{equation}
\end{subequations}
where $N_e$ is the number of electronic states (for a single spin/single lead) and $N_b$ is the number of
bath modes in the simulation.  In this work, we choose a constant $\rho_e(E)$, i.e., an equidistant
discretization of the interval $[-2\beta_e, 2\beta_e]$, to discretize the electronic continuum.  For 
the vibrational bath, $\rho(\omega)$ is chosen as
\begin{equation}
        \rho(\omega) = \frac{N_b+1}{ \omega_c} e^{-\omega/\omega_c}.
\end{equation} 
Within a given time scale the numbers of electronic states and bath 
modes are systematically increased to reach converged results for the quantum dynamics in the 
extended condensed phase system.

\subsection{Initial state and calculation of current}

In this paper the observable of interest for studying transport through molecular junctions is the 
current for a given source-drain bias voltage, given by (we use atomic units where $\hbar = e = 1$)
\begin{subequations}
\begin{equation}
	I_L(t) = - \frac{d N_L(t)} {dt} = -\frac{1}{{\rm tr}[\hat{\rho}]} {\rm tr}
        \left\{ \hat{\rho} e^{i\hat{H}t} i[\hat{H}, \hat{N}_{L}] e^{-i\hat{H}t} \right\},
\end{equation}
\begin{equation}
	I_R(t) = \frac{d N_R(t)} {dt} = \frac{1}{{\rm tr}[\hat{\rho}]} {\rm tr}
        \left\{ \hat{\rho} e^{i\hat{H}t} i[\hat{H}, \hat{N}_{R}] e^{-i\hat{H}t} \right\}.
\end{equation}
\end{subequations}
Here, $N_{L/R}(t)$ denotes the  time-dependent charge in each lead, defined as 
\begin{equation}
	N_{\zeta}(t) = \frac{1}{{\rm tr}[\hat{\rho}]} {\rm tr}
        [\hat{\rho} e^{i\hat{H}t} \hat{N}_{\zeta} e^{-i\hat{H}t} ], 
\end{equation}
and  $\hat{N}_{\zeta} = \sum_{k_\zeta} c_{k_\zeta}^+ c_{k_\zeta}$
is the occupation number operator for the electrons in each lead ($\zeta=L, R$).
For Hamiltonian (\ref{Htot}) the explicit 
expression for the current operator is given as
\begin{equation}\label{Iop}
	\hat{I}_\zeta \equiv i[\hat{H}, \hat{N}_{\zeta}] = i \sum_{k_\zeta}
	V_{dk_\zeta} ( d^+ c_{k_\zeta} - c_{k_\zeta}^+ d ), \;\;\; 
	\zeta=L, R. 
\end{equation}
In the expressions above, $\hat{\rho}$ denotes the 
initial density matrix representing a grand-canonical ensemble for each lead and a certain
preparation for the bridge state
\begin{subequations}\label{Initden}
\begin{equation}
	\hat{\rho} = \hat{\rho}_d^0 \;{\rm exp} \left[ -\beta (\hat{H}_0 
          - \mu_L \hat{N}_L - \mu_R \hat{N}_R) \right],
\end{equation}
\begin{equation}
	\hat{H}_0 = \sum_{k_L,\sigma} E_{k_L} \hat{n}_{k_L,\sigma}
	+ \sum_{k_R,\sigma} E_{k_R} \hat{n}_{k_R,\sigma}  + \hat{H}_{\rm nuc}^0.
\end{equation}
\end{subequations}
Here $\hat{\rho}_d^0$ is the initial reduced density matrix for the bridge state, which is usually chosen as
a pure state representing an occupied or an empty bridge state, and $\hat{H}_{\rm nuc}^0$ defines the initial
bath equilibrium distribution. 

Various initial states can be considered.  For example, one may choose an initially unoccupied bridge 
state and the nuclear degrees of freedom equilibrated with this state, i.e. an  unshifted 
bath of oscillators with
\begin{equation}
{\hat H}_{\rm nuc}^{0} = \frac{1}{2} \sum_j \left[ P_j^2 + \omega_j^2  Q_j^2 \right].
\end{equation}
On the other hand, one may also start with a fully occupied bridge state and 
a bath of oscillators in equilibrium with the occupied bridge state
\begin{equation}
{\hat H}_{\rm nuc}^{0'} = \frac{1}{2} \sum_j \left[ P_j^2 + \omega_j^2 \left(Q_j + 2
\frac{c_j}{\omega_j^2}\right)^2 \right].
\end{equation}
Other initial states may also be prepared. The initial state may affect the
transient dynamics profoundly. The dependence of the steady-state current on
the initial density matrix is a more complex issue. Recent investigations for a model without
electron-electron interaction seem to indicate that different
initial states may lead to different (quasi)steady states,\cite{Gogolin02,Galperin05,Albrecht12} 
although this has been debated.\cite{Alexandrov07}  
In this paper, we limit
our discussion to physical regimes where the steady state current does not depend on the initial
state.  When comparing the time-dependent current from the scattering state representation with
that from the original Hamiltonian discussed above,  we typically choose initial conditions that 
are close to the final steady state, e.g., an unoccupied initial bridge state if its energy is 
higher than the Fermi level of the leads and an occupied bridge state otherwise. We also use the
average current 
\begin{equation}
	I(t) = \frac{1}{2} [ I_R(t) + I_L(t) ],
\end{equation} 
for such a purpose, which minimizes transient effects.

In our simulations the continuous set of electronic states of the leads is represented by 
a finite number of states.  The number of states required to properly describe the 
continuum limit depends on the time $t$. The situation is thus similar to that of a quantum reactive 
scattering calculation in the presence of a scattering continuum, where, with a finite number of basis 
functions, an appropriate absorbing boundary condition is added to mimic the correct outgoing 
Green's function.\cite{Goldberg1978,Kosloff1986,Neuhauser1989,Seideman1991} 
Employing the same strategy for the present problem, the 
regularized electric current is given by
\begin{equation}
	I^{\rm reg}  = \lim_{\eta \to 0^+} \int_0^{\infty} dt \, \frac{dI(t)}{dt} \, e^{-\eta t}.
\end{equation}

The regularization parameter $\eta$ is similar (though not identical) to the formal convergence parameter
in the definition of  the Green's function in terms of the time evolution operator
\begin{equation}
	G(E^+) = \lim_{\eta \to 0^+} (-i) \int_0^{\infty} dt\, e^{i(E+i\eta-H)t}.
\end{equation}
In numerical calculations, $\eta$ is chosen in a similar way as the absorbing potential used in quantum
scattering calculations.\cite{Goldberg1978,Kosloff1986,Neuhauser1989,Seideman1991} In particular, 
the parameter $\eta$ has to be large enough to accelerate the convergence but still sufficiently small
in order not to affect the correct result.  While in the reactive scattering calculation
$\eta$ is often chosen to be coordinate dependent, in our simulation $\eta$ is chosen
to be time dependent
\begin{equation}\label{damping}
\eta(t) = \left\{
              \begin{array}{ll}
                   0 & \quad (t<\tau)\\
                   \eta_0\cdot (t-\tau)/t & \quad (t>\tau) .
              \end{array}
       \right.
\end{equation}
Here $\eta_0$ is a damping constant, $\tau$ is a cutoff time beyond which a steady state charge
flow is approximately reached.  As the number of electronic states increases, one may choose a 
weaker damping strength $\eta_0$ and/or longer cutoff time $\tau$.  The former approaches zero and the
latter approaches infinity for an infinite number of states.  
For the systems considered in this work, convergence can be reached with a typical $\tau =$ 30-80 fs 
(a smaller $\tau$ for less number of states) and $1/\eta_0 =$ 3-10 fs.

\section{Scattering State Representation of the Model}\label{modelscatter}

In previous work \cite{wan09:024114,Wang11} we have used the above Hamiltonian
with the ML-MCTDH-SQR  theory for studying vibrationally coupled electron transport
through molecular junctions.  It provided important benchmark results to guide further
development of the approximate theories. However, it was also found that upon increasing the bias voltage
the number of configurations required in the ML-MCTDH representation of the wave
function to achieve convergence becomes very large.  This is
so even without including the contribution from the nuclear bath,
Eq.~(\ref{Hel-nuc}), i.e.\ for the so-called noninteracting model.  In fact,
in many cases, including the vibrational coupling accelerates the convergence of the ML-MCTDH-SQR
simulation within a given time scale (see below for a physical explanation).

Such a performance is not surprising if one considers the electronic Hamiltonian in Eq.~(\ref{Hel}).
Similar to the example of a quadratic Hamiltonian with full Hessian, all the electronic degrees
of freedom in Eq.~(\ref{Hel}) may be strongly coupled, thus creating significant artificial 
correlations in the quantum dynamics of the model.  Including the vibronic contribution in 
Eq.~(\ref{Hel-nuc}) may suppress such artificial correlations, though it also introduces vibrationally 
induced `true' electronic correlation effect.  Within the current form of the Hamiltonian, these
artificial correlations are real and cannot be removed in a variational calculation.

To circumvent this problem and to reduce the artificial correlation, we take advantage of the fact that
the Green's function and also the steady
state current for the noninteracting model defined by Eq.~(\ref{Hel}) can be
obtained analytically. Thus, the electronic Hamiltonian Eq.~(\ref{Hel}) can be
represented in the basis of
its eigenstates, which are scattering states. The resulting scattering states
representation of the model has been used before to describe quantum transport
in combination with different methods.\cite{Hershfield93,Doyon06,Han07,Oguri07,Anders08}

\subsection{General formulation}\label{naive}

In the scattering state representation
\cite{Hershfield93,Doyon06,Han07,Oguri07,Anders08}  a set of scattering state
operators  are introduced to diagonalize the Hamiltonian (\ref{Hel}).      
These operators $\gamma_{k\zeta}^+$ satisfy the 
Lippmann-Schwinger equation ($\zeta = L, R$)
\begin{subequations}
\begin{equation}
	\gamma_{k\zeta}^+ = c_{k\zeta}^+ + \frac{1}{E_{k\zeta} + i\eta - L_0} [\hat V, \gamma_{k\zeta}^+],
\end{equation}
or the equivalent Heisenberg equation
\begin{equation}
	[\hat H_{\rm el}, \gamma_{k\zeta}^+] = E_{k\zeta} \gamma_{k\zeta}^+
					+ i\eta (\gamma_{k\zeta}^+ -  c_{k\zeta}^+).
\end{equation}
\end{subequations}
Here $\hat H_{\rm el} = \hat H_0 + \hat V$, $\hat L_0 = [\hat H_0,]$ is a Liouville operator, $\eta$ is an infinitesimal, and
\begin{subequations}
\begin{equation}
	 \hat H_0 = E_d d^+ d + \sum_{k,\zeta} E_{k\zeta} c_{k\zeta}^+ c_{k\zeta},
\end{equation}
\begin{equation}
	\hat V = \sum_{k,\zeta} V_{dk\zeta} ( d^+ c_{k\zeta} + c_{k\zeta}^+ d ).
\end{equation}
\end{subequations}
The scattering state operators $\gamma_{k\zeta}^+/\gamma_{k\zeta}$ fulfill the usual fermionic commutation
relations. They  can be expanded in terms of $d^+$ and $c_{k_\zeta}^+$,
and the formal solution to the Lippmann-Schwinger equation is given by
\begin{equation}\label{forward}
	\gamma_{k\zeta}^+ = c_{k\zeta}^+ + V_{dk\zeta} g_d^0 (E_{k\zeta})
	\left[ d^+ + \sum_{k',\zeta'}\frac{V_{dk'{\zeta'}}}{E_{k\zeta}+i\eta - E_{k'\zeta'}}
	c_{k'\zeta'}^+ \right],
\end{equation}
where $g_d^0 (E_{k\zeta})$ is the retarded Green's function for the bridge state
\begin{subequations}
\begin{equation}
	g_d^0 (E_{k\zeta}) = \frac{1}{E_{k\zeta}+i\eta-E_d-\Sigma(E_{k\zeta})},
\end{equation}
\begin{equation}
	\Sigma(E_{k\zeta}) = \sum_{k',\zeta'} \frac{|V_{dk'{\zeta'}}|^2}{E_{k\zeta}+i\eta - E_{k'\zeta'}}.
\end{equation}
\end{subequations}
For the tight-binding parameterization, Eq.~(\ref{tight}), the self energy 
$\Sigma(E)= \Sigma_L(E) + \Sigma_R(E)$ has the following analytic form
\begin{equation}
	\Sigma_{\zeta}(E) = -\frac{i}{2} \Gamma_{\zeta}(E) + \Delta_{\zeta}(E), \hspace{1cm}
	\zeta = L, R,
\end{equation}
where $\Gamma_{\zeta}(E)$ is given by Eq.~(\ref{tight}), and
\begin{equation}
        \Delta_\zeta (E) = \left\{ \begin{array}{ll} \frac{\alpha_e^2}{2\beta_e^2} (E - \mu_\zeta) 
		\hspace{1cm} & |E-\mu_\zeta| \leq 2 \beta_e \\
                \frac{\alpha_e^2}{2\beta_e^2} \left[(E - \mu_\zeta) \mp \sqrt{4\beta_e^2-(E-\mu_\zeta)^2} 
		\right] \hspace{1cm} &  \pm (E-\mu_\zeta) > 2 \beta_e \end{array}  \right.,
\end{equation}

In the following, we assume that the purely electronic problem defined by the
Hamiltonian Eq.~(\ref{Hel}) does not possess bound states, so that the set of the scattering states
provide a complete basis. As a result, the electronic Hamiltonian is
diagonal in the scattering representation and has the form
\begin{equation}
	\hat H_{\rm el} = \sum_{k,\zeta} E_{k\zeta} \gamma_{k\zeta}^+ \gamma_{k\zeta}
	       = \sum_{k_L} E_{k_L} \gamma_{k_L}^+ \gamma_{k_L} + \sum_{k_R} E_{k_R} \gamma_{k_R}^+ \gamma_{k_R}.
\end{equation}
For the vibronic part of the Hamiltonian in Eq.~(\ref{Hel-nuc}) and the current operator in Eq.~(\ref{Iop}), 
one applies the inverse relation of Eq.~(\ref{forward})
\begin{subequations}\label{backward}
\begin{equation}
	d^+ =  \sum_{k,\zeta} \gamma_{k\zeta}^+ \left[ g_d^0 (E_{k\zeta}) V_{dk\zeta}  \right]^*
\end{equation}
and
\begin{equation}
	c_{k\zeta}^+ = \gamma_{k\zeta}^+ +  \sum_{k',\zeta'} \gamma_{k'\zeta'}^+ 
	\left[ g_d^0 (E_{k'\zeta'}) \frac{ V_{dk'\zeta'} V_{dk\zeta}^* } {E_{k'\zeta'} +i\eta - E_{k\zeta}}  
	\right]^*,
\end{equation}
\end{subequations}
which, when substituting into Eq.~(\ref{Hel-nuc}) and Eq.~(\ref{Iop}), yields the necessary expressions for
implementation.  It is noted that both the number operator $d^+d$ and the current operator $\hat{I}_\zeta$
are more complex in the scattering state representation.  This is the price for reducing the 
artificial correlation effect in the quantum dynamics simulation.

Finally, the initial density matrix, which will be used in the current
simulation based on the scattering state representation, is given by
\begin{equation}\label{scatt_des}
	\hat{\rho} = \frac{ e^{-\beta({\hat H}_1 - \hat Y)}}{{\rm tr}[e^{-\beta({\hat H}_1 - \hat Y)}]},
\end{equation}
where $H_1$ is the Hamiltonian without the vibronic coupling
\begin{eqnarray}
	{\hat H_1} &=& \sum_{k_L} E_{k_L} \gamma_{k_L}^+ \gamma_{k_L}  
                  + \sum_{k_R} E_{k_R} \gamma_{k_R}^+ \gamma_{k_R}
		+ {\hat H}_{\rm nuc}^{0} \\
 &=& {\hat H}_{\rm el} + {\hat H}_{\rm nuc}^{0} \nonumber
\end{eqnarray}
and Y is the bias operator
\begin{equation}
	\hat Y = \mu_L \sum_{k_L} \gamma_{k_L}^+ \gamma_{k_L} + \mu_R \sum_{k_R} \gamma_{k_R}^+ \gamma_{k_R}.
\end{equation}

The density matrix in Eq.\ (\ref{scatt_des}) can be factorized
\begin{equation}
	\hat{\rho} = \hat{\rho}_{\rm el}\hat{\rho}_{\rm nucl},
\end{equation}
into an electronic part
\begin{equation}
	\hat{\rho}_{\rm el} = \frac{ e^{-\beta({\hat H}_{\rm el} - \hat Y)}}{{\rm tr}[e^{-\beta({\hat H}_{\rm el} - \hat Y)}]},
\end{equation}
and a nuclear part
\begin{equation}
	\hat{\rho}_{\rm nucl} = \frac{ e^{-\beta {\hat H}_{\rm nuc}^{0}}}{{\rm
            tr}[e^{-\beta {\hat H}_{\rm nuc}^{0}}]}.
\end{equation}
The electronic density matrix $\hat{\rho}_{\rm el}$ describes the steady state
of  the purely electronic model
(without electron-vibrational coupling), i.e.
\begin{equation}
	\lim_{t \to \infty}{\rm tr} \{\rho_0  e^{i\hat{H}_{\rm el}t} \hat A
        e^{-i\hat{H}_{\rm el}t} \} =
        {\rm tr} \left\{ \frac{ e^{-\beta(\hat{H}_{\rm el} - \hat Y)}}{{\rm tr}
       [e^{-\beta(\hat{H}_{\rm el} -
              \hat Y)}]} \, \hat A \right \},
\end{equation}
holds for a given density matrix $\rho_0$ and electronic operator
$\hat A$.\cite{Hershfield93,Gelin09} In particular, the steady state current for vanishing
electronic-vibrational coupling is given by
\begin{equation}
	\lim_{t \to \infty} I_{L/R} (t) = \mp
        {\rm tr} \left\{ \frac{ e^{-\beta(\hat{H}_{\rm el} - \hat Y)}}{{\rm tr}[e^{-\beta(\hat{H}_{\rm el} -
              \hat Y)}]} \, \hat  I_{L/R}\ \right \},
\end{equation}
Assuming  a unique steady state, both initial states, Eqs.\ (\ref{scatt_des}) and
(\ref{Initden}),  will result in the same steady-state current, however, the transient
dynamics will be different.

\subsection{Optimization of the transport calculation including vibronic coupling}\label{specific}

It is straightforward to implement the above Hamiltonian in the scattering state representation
for the ML-MCTDH-SQR simulation of the current.  Without vibronic coupling,
the model corresponds to a
non-interacting system.  We have verified that with a sufficient number of
states used in the discretization of the electronic continua of the leads, the steady
state current agrees with the Landauer formula, which is exact in this
noninteracting case.  Moreover, the ML-MCTDH-SQR only needs one configuration
to obtain the numerically exact result, as it should be. 

Including vibronic coupling, this is no longer the case because the systems represents
a true many-body problem. In this case, we have
found that the direct application of the scattering state representation as
introduced above may result in serious numerical problems except for a range of
positive energies for the original bridge state ($E_d > E_f$) where limited success is achieved. 
In particular, when $E_d$
approaches zero or becomes negative, the simulated steady-state current sometimes
becomes negative even with a relatively large number of scattering states and configurations in the
ML-MCTDH-SQR calculation, which is clearly unphysical.  

This numerical problem may be understood from the technical aspects of applying a finite basis 
set representation to study dynamics.  In principle, an infinite basis expansion 
can solve the problem exactly, but in practice one wants to keep the number of basis
functions as small as possible to make the simulation feasible.  To achieve this, it is advantageous to 
keep the (complete) basis sets as close as possible to the eigenfunctions of the overall Hamiltonian.  
If this is not done appropriately, one may need a prohibitively large number of basis functions 
(i.e., an ``infinite'' basis in a practical sense) to obtain numerical convergence, such as what
we have observed here.  In the discretized scattering state representation, the overall current 
comes from the contribution of all the scattering channels. The weight of each channel/scattering state 
is determined properly from the solution to the pure electronic Hamiltonian.  However, when including the
vibrational degrees of freedom in the Hamiltonian and electron-vibrational coupling, 
such a basis set may be far from optimal.  
The channels that contribute the most to the original, non-interacting electronic system may not 
be important at all to the overall true many-body transport problem. On the other hand, other channels
that were insignificant previously may now become important.  

According to our investigations,
this numerical problem can be circumvented by noticing the fact that the scattering
state representation is not unique. In particular, there is an arbitrariness in
defining the bridge state energy, e.g., one can rewrite  Eq.~(\ref{Htot}) as
\begin{subequations}
\begin{eqnarray}
	\hat H_{\rm el} &=& (E_d - \Delta E) d^+ d + \sum_{k_L} E_{k_L} c_{k_L}^+ c_{k_L}
	+ \sum_{k_R} E_{k_R} c_{k_R}^+ c_{k_R} \\ \nonumber
  &&	+ \sum_{k_L} V_{dk_L} ( d^+ c_{k_L} + c_{k_L}^+ d )
	+ \sum_{k_R} V_{dk_R} ( d^+ c_{k_R} + c_{k_R}^+ d ),
\end{eqnarray}
\begin{equation}
	\hat H_{\rm el-nuc} = \frac{1}{2} \sum_j ( P_j^2 + \omega_j^2 Q_j^2 ) 
	+ d^+ d (\Delta E+  \sum_j 2 c_j Q_j).
\end{equation}
\end{subequations}
With a different choice of $\Delta E$, the scattering state operators are different and so will be 
the performance of the vibronic transport calculation.  Our numerical tests show
that choosing specifically the value of $\Delta E$, which is obtained by
completing the square in the vibronic part, i.e.
\begin{subequations}
\begin{eqnarray}\label{opt_Hel}
	\hat H_{\rm el} &=& \left(E_d - \sum_j \frac{2c_j^2}{\omega_j^2} \right) d^+ d 
	+ \sum_{k_L} E_{k_L} c_{k_L}^+ c_{k_L}
	+ \sum_{k_R} E_{k_R} c_{k_R}^+ c_{k_R} \\ \nonumber
  &&	+ \sum_{k_L} V_{dk_L} ( d^+ c_{k_L} + c_{k_L}^+ d )
	+ \sum_{k_R} V_{dk_R} ( d^+ c_{k_R} + c_{k_R}^+ d ),
\end{eqnarray}
\begin{equation}
	\hat H_{\rm el-nuc} = \frac{1}{2} \sum_j \left[ P_j^2 + 
	\omega_j^2 \left( Q_j + d^+ d \frac{2c_j}{\omega_j^2} \right)^2  \right],
\end{equation}
\end{subequations}
corresponding to $\Delta E=\sum_j 2 c_j^2/\omega_j^2$, 
results in  a particularly good performance of the transport calculations.
It is recognized that for this choice of $\Delta E$, the correction to the
bridge state  energy, $\sum_j 2 c_j^2/\omega_j^2$, is the
reorganization energy (polaron shift) for electron transfer between the
bridge and the lead states and thus this selection also has a clear  physical meaning.

With the choice of the scattering state operators that diagonalize the 
$H_{\rm  el}$ in Eq.\ \ref{opt_Hel}, the effective
Hamiltonian for the overall vibronic model performs well in all regimes we have tested.  The numerical
examples presented below will illustrate this point.

\section{Simulation of transport in the scattering state
  representation using the  ML-MCTDH-SQR method}\label{results}

To simulate transport in the model described above, we employ the Multilayer
Multiconfiguration  Time-Dependent Hartree Theory in Second Quantization
Representation  (ML-MCTDH-SQR),\cite{wan10:78} which
allows a numerically exact treatment of the many-body problem. The method has
been described in detail elsewhere.\cite{wan10:78} It is based on the ML-MCTDH
theory\cite{wan03:1289}, which is a rigorous variational method to propagate wave packets 
in complex systems with many degrees of freedom.
In the ML-MCTDH theory\cite{wan03:1289} the wave function is represented by a recursive, layered expansion, 
\begin{subequations}\label{psiml}
\begin{equation}\label{L1}
        |\Psi (t) \rangle = \sum_{j_1} \sum_{j_2} ... \sum_{j_p}
        A_{j_1j_2...j_p}(t) \prod_{\kappa=1}^{p}  |\varphi_{j_\kappa}^{(\kappa)} (t) \rangle,
\end{equation}
\begin{equation}\label{L2}
        |\varphi_{j_\kappa}^{(\kappa)}(t)\rangle =  \sum_{i_1} \sum_{i_2} ... \sum_{i_{Q(\kappa)}}
        B_{i_1i_2...i_{Q(\kappa)}}^{\kappa,j_\kappa}(t) \prod_{q=1}^{Q(\kappa)}  
	|v_{i_q}^{(\kappa,q)}(t) \rangle,
\end{equation}
\begin{equation}\label{L3}
        |v_{i_q}^{(\kappa,q)}(t)\rangle  = \sum_{\alpha_1} \sum_{\alpha_2} ... 
	\sum_{\alpha_{M(\kappa,q)}}
        C_{\alpha_1\alpha_2...\alpha_{M(\kappa,q)}}^{\kappa,q,i_q}(t) 
	\prod_{\gamma=1}^{M(\kappa,q)}  
	|\xi_{\alpha_\gamma}^{\kappa,q,\gamma}(t) \rangle,
\end{equation}
\begin{equation}
	... \nonumber
\end{equation}
\end{subequations}
where $A_{j_1j_2...j_p}(t)$, $B_{i_1i_2...i_{Q(\kappa)}}^{\kappa,j_\kappa}(t)$,
$C_{\alpha_1\alpha_2...\alpha_{M(\kappa,q)}}^{\kappa,q,i_q}(t)$ and so on are the
expansion coefficients for the first, second, third, ..., layers, respectively;
$|\varphi_{j_\kappa}^{(\kappa)} (t) \rangle$, $|v_{i_q}^{(\kappa,q)}(t) \rangle$,
$|\xi_{\alpha_\gamma}^{\kappa,q,\gamma}(t) \rangle$, ..., are the ``single particle'' 
functions (SPFs) for the first, second, third, ..., layers.  
In Eq.~(\ref{L1}), $p$ denotes the number of single
particle (SP) groups/subspaces for the first layer.  Similarly, $Q(\kappa)$ in Eq.~(\ref{L2})
is the number of SP groups for the second layer that belongs to the $\kappa$th SP
group in the first layer,  i.e., there are a total of $\sum_{\kappa=1}^{p} Q(\kappa)$
second layer SP groups.  Continuing along the multilayer hierarchy, 
$M(\kappa,q)$ in Eq.~(\ref{L3}) is the number of SP groups for the third layer that belongs 
to the $q$th SP group of the second layer and the $\kappa$th SP group of the first layer,  
resulting in a total of $\sum_{\kappa=1}^{p} \sum_{q=1}^{Q(\kappa)} M(\kappa,q)$ third 
layer SP groups.

Each summation in the expressions above is over the number of SPFs in that single particle 
group and the total combination defines the size of the configurational space in that layer,
e.g., if there are 4 SP groups and we use 10 SPFs per group, the configurational space
is $10^4 = 10000$. 
If there is no correlation at all between different degrees of freedom, only one SPF is
required for each group, resulting in one overall configuration.  As the correlation becomes
stronger, more SPFs are required to achieve convergence, which results in a larger 
configuration space.  Thus, the number of SPFs is a good indicator of the correlation 
strength and will be used for analysis purposes below.

\section{Results and Discussion}

In this section, we apply the methodology discussed above to study
vibrationally-coupled charge transport.  In particular, we discuss for selected
examples the performance of the simulation in the scattering-state representation as
compared to the original representation.  Without specifying otherwise, we will use the tight-binding
parameterization for the electronic part of the model, as described in Sec.~\ref{modeltight} and used 
equivalently in the scattering state representation, Sec.~\ref{modelscatter}.  The tight-binding 
parameters for the function $\Gamma (E)$ are $\alpha_e = 0.2$ eV, $\beta_e = 1$ eV, corresponding 
to a moderate molecule-lead coupling an a bandwidth of 4 eV.

\subsection{Transport without vibronic coupling}

If the Hamiltonian only contains the pure electronic part, i.e.\  without
electronic vibrational coupling, it corresponds to the
non-interacting transport model.  Employing the scattering-state
representation, in this case one SPF for each SP group, i.e., one
overall configuration representing the wave function, gives the numerically exact 
steady-state current.  This was verified by comparing the numerical results obtained 
with one configuration of the wave function to the Landauer formula.  In
contrast, for the original
representation of the model described in Sec.~\ref{modeltight}, there is significant artificial 
correlation, especially for a large source-drain bias voltage, such that a large
configurational space is required to achieve convergence.

\subsection{Vibrationally coupled electron transport for $E_d > E_f$}\label{large_Ed}

\begin{figure}[!ht]
\begin{flushleft}
(a)
\end{flushleft}
\includegraphics[clip,width=0.45\textwidth]{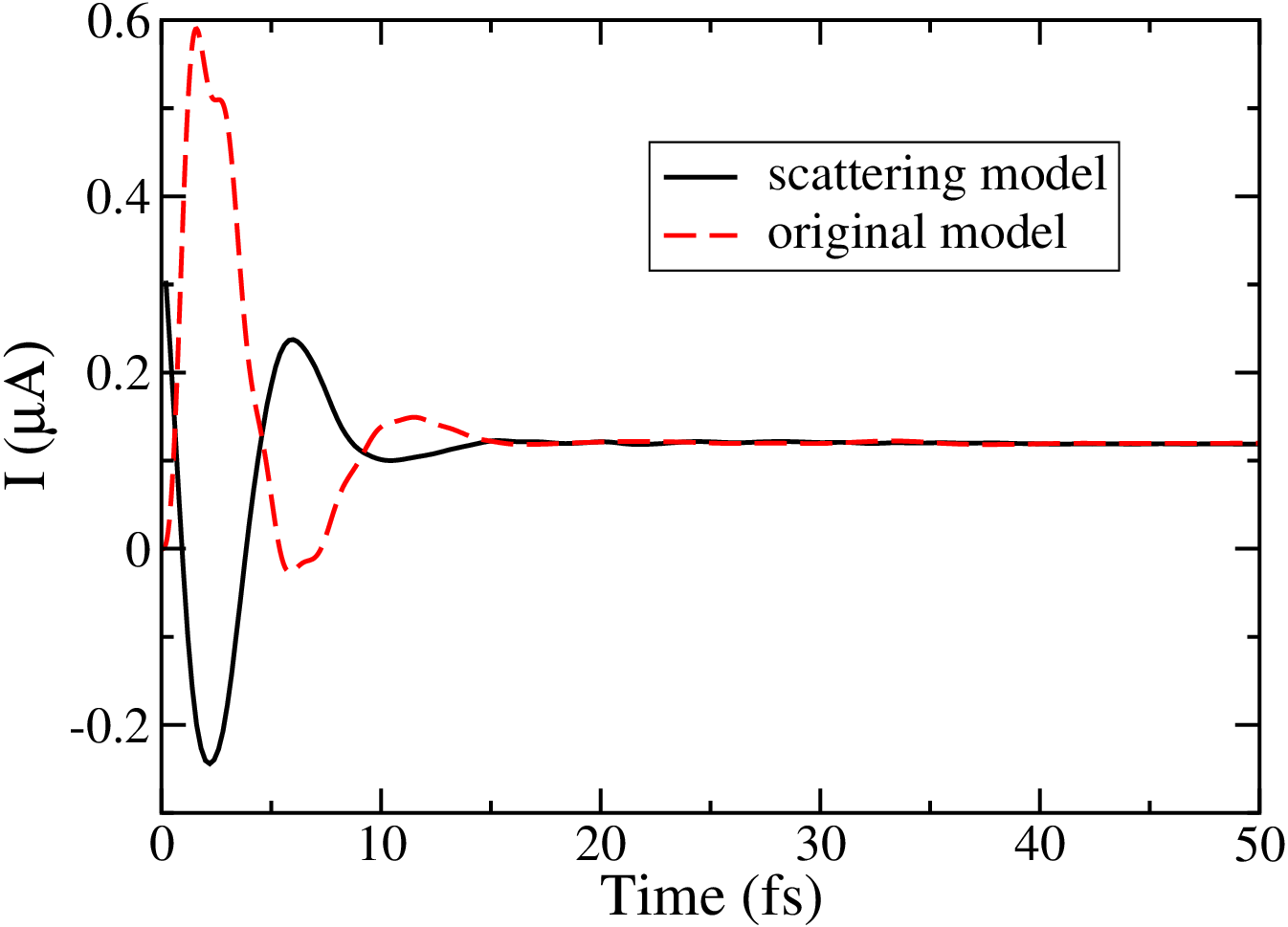}

\begin{flushleft}
(b)
\end{flushleft}
\includegraphics[clip,width=0.45\textwidth]{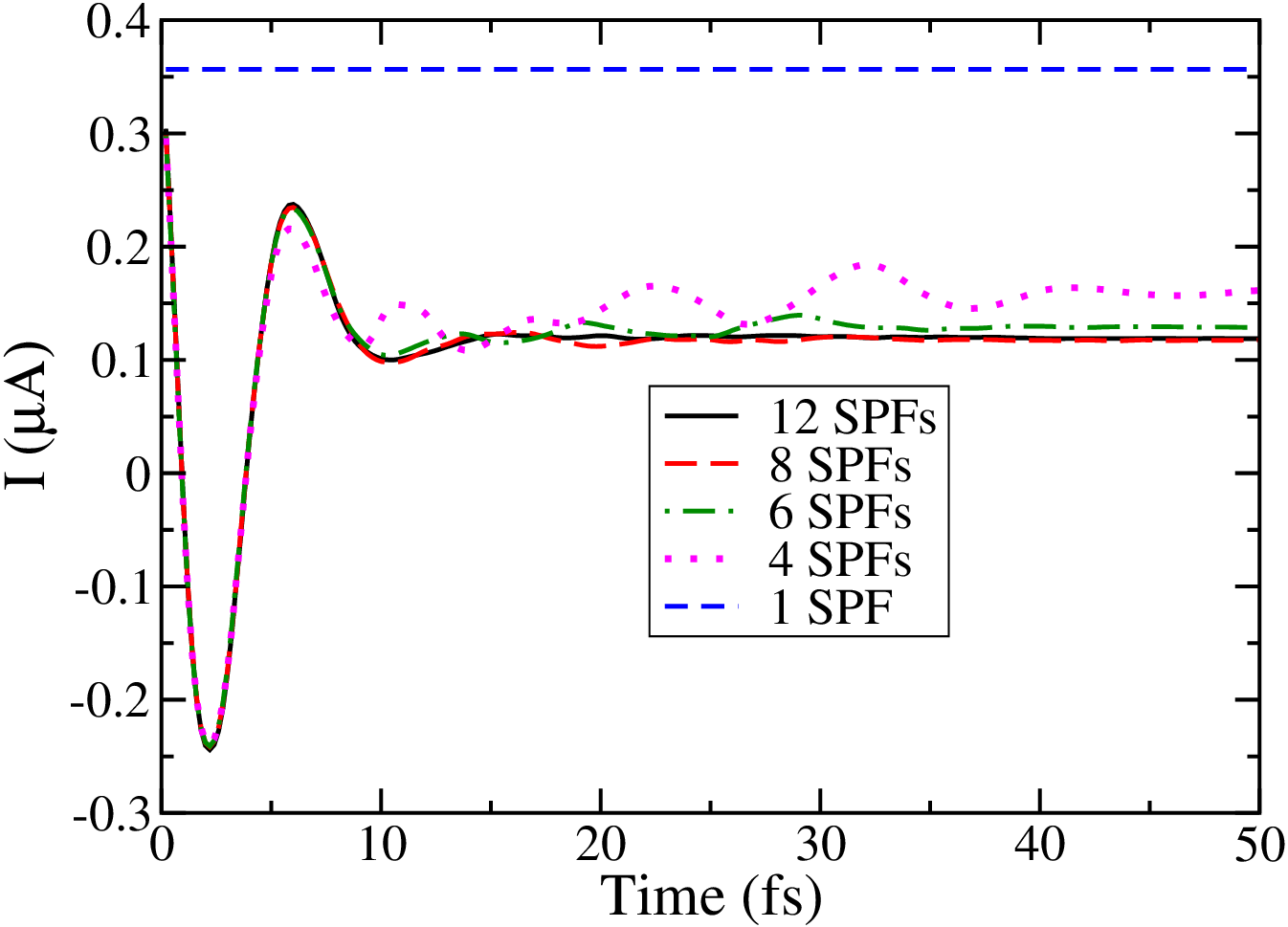}

\caption{(a) Time-dependent current $I(t)$ for different representations of
  the model Hamiltonian (scattering state vs. original representation),
   (b) Convergence of $I(t)$ versus the number of SPFs for the electronic part, where
6 SPFs are used for each nuclear SP group. 
The model parameters are: $\alpha_e = 0.2$eV, $\beta_e = 1$eV, and $E_d - E_f = 0.5$eV
for the electronic part; $\lambda = 2000{\rm cm}^{-1}$ and $\omega_c = 500{\rm cm}^{-1}$
for the vibrational bath. The source-drain voltage is $V = 0.1$V.  }
\label{fig1}
\end{figure}

Including electronic-vibrational coupling, the model becomes a many-body
problem with ``true'' correlation.
We first consider a physical regime where, without vibronic coupling, the electronic energy 
of the bridge state is located above the Fermi level of the leads, $E_d - E_f
= 0.5$eV.  For small bias voltages, this corresponds to the non-resonant
transport regime.
Fig.~\ref{fig1}a shows the converged time-dependent currents obtained with the original
model Hamiltonian in Sec.~\ref{modeltight} and with the scattering state Hamiltonian.
Thereby, the characteristic frequency of the bath has been chosen as $\omega_c = 500$~cm$^{-1}$ and the 
overall electronic-nuclear coupling strength is determined by the reorganization energy, 
$\lambda = 2\alpha\omega_c = 2000$~cm$^{-1}$.  The source-drain bias voltage is 0.1V.  Despite
the difference in the transient behavior of $I(t)$, which is expected due to the difference
in the initial states between the two model representations, the stationary values of the
current agree within 1\% relative difference.  This small numerical error is remarkable considering the two
drastically different representations of the model and the wave function.  Fig.~\ref{fig1}b shows
the convergence behavior of $I(t)$ versus the number of SPFs for each electronic SP group,
where 6 SPFs for each nuclear SP group provides a converged solution.  It is
seen that the result obtained with
8 SPFs per each electronic SP group agrees perfectly with a larger 12 SPFs simulation.  Even
6 SPFs per electronic SP group gives a result that is within 10\% relative error of the converged
answer.  On the other hand, to achieve convergence for the original model requires 40
SPFs per electronic SP group to generate the result in Fig.~\ref{fig1}a.

This shows that by using the Hamiltonian in the scattering state representation
much of the artificial correlation has been removed.  This results in a significantly smaller
configuration space than that required for the original model Hamiltonian.  A comparison of 
results obtained with different number of SPFs also reveals the extent of the ``true''
correlation effect, which is helpful for analyzing the physics of the problem. 

It is also interesting to consider the result obtained with 1 SPF, i.e., only one overall configuration.
Within the scattering state representation, this corresponds to the
Hartree-Fock approximation for the present vibronic model.  Fig.~\ref{fig1}b shows
that in this case: (i) there is no transient dynamics in the simulated $I(t)$ and (ii) the stationary
current is approximately three times as large as the true result.  Further inspection shows that the
one-configuration result agrees with a pure electronic model (Landauer formula), where the
bridge state energy is shifted by the bath reorganization energy, i.e., $E_d' = E_d - \lambda$.  For
this particular example, this brings the level closer to the chemical potential of the electrodes. 
As a result, the current is enhanced compared with that without vibronic coupling.  Although the
one-configuration result captures this effect qualitatively, it is not
quantitative at all, being a factor of
three too large.

\begin{figure}[!ht]
\begin{flushleft}
(a)
\end{flushleft}
\includegraphics[clip,width=0.45\textwidth]{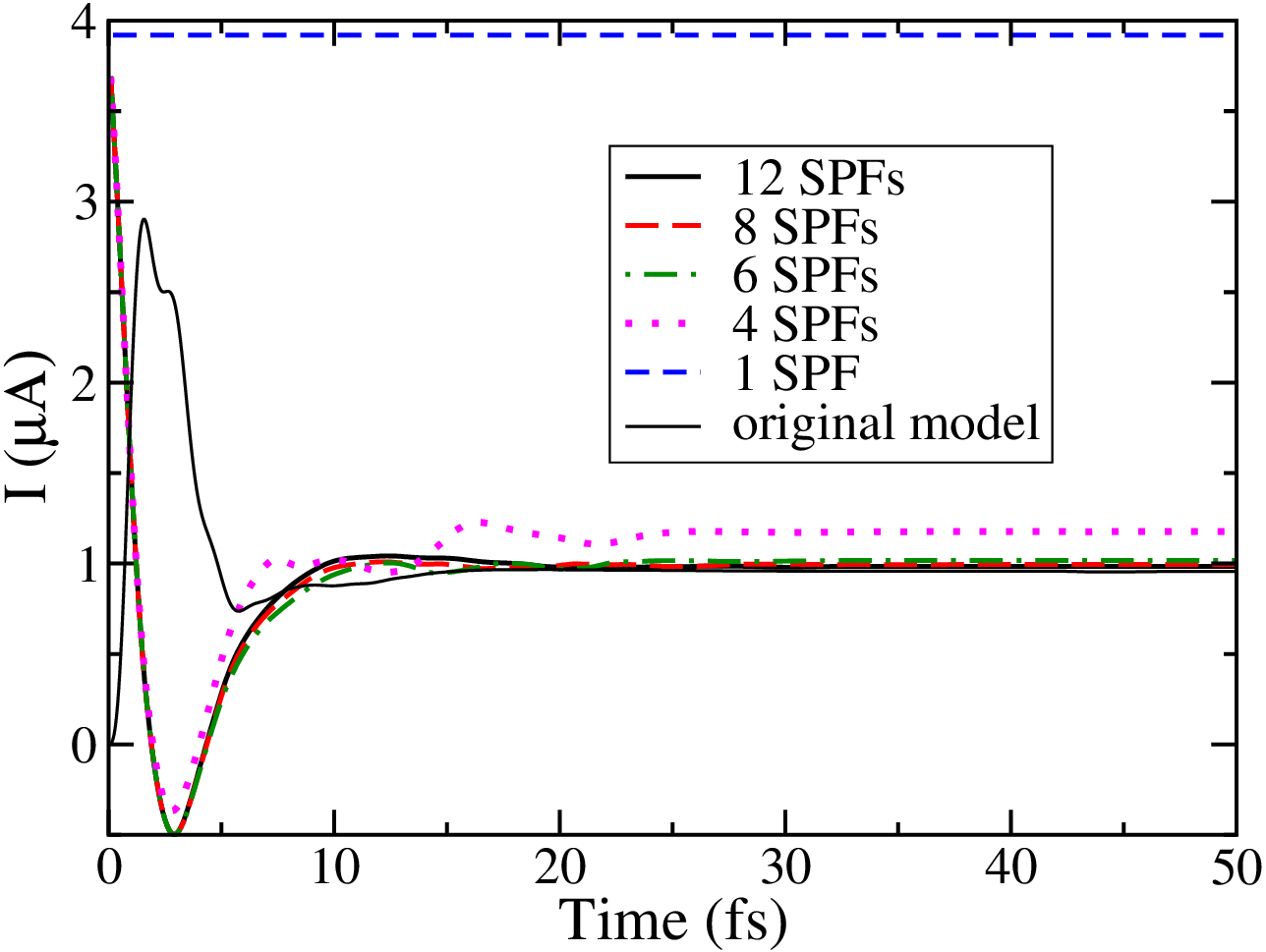}

\begin{flushleft}
(b)
\end{flushleft}
\includegraphics[clip,width=0.45\textwidth]{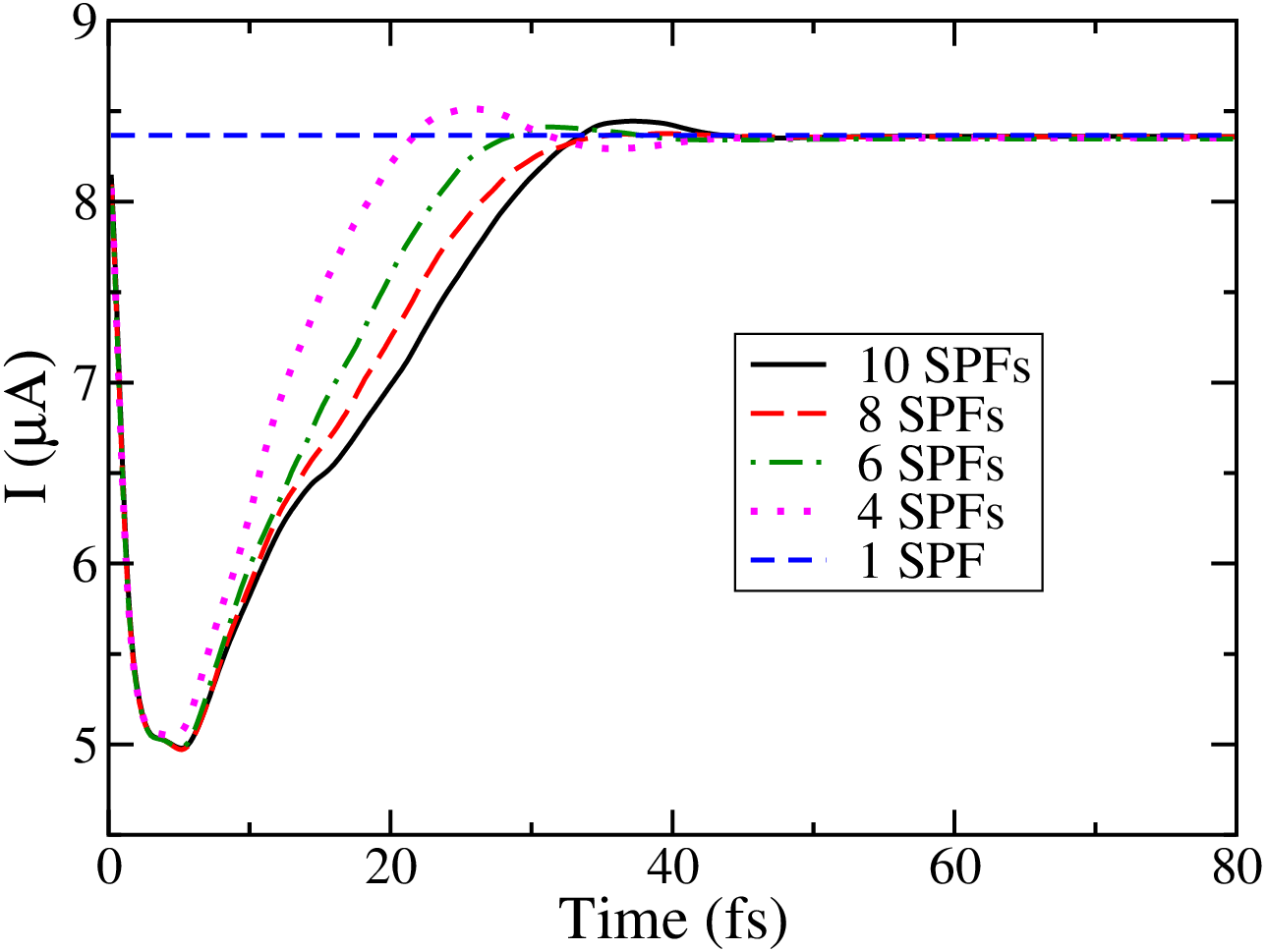}

\caption{Convergence of $I(t)$ versus the number of SPFs.  The parameters are the same as Fig.~\ref{fig1} 
except for the bias voltage: (a) $V = 0.5$V, (b) $V = 1$V.}
\label{fig2}
\end{figure}

Figure~\ref{fig2} shows the time-dependent current for larger bias
voltages, 0.5V and 1V, where all other parameters are the same as in
Fig.~\ref{fig1}.  For $V=0.5$V, as depicted in Fig.~\ref{fig2}a, about
6-8 SPFs per electronic SP group gives the converged result that is in good agreement with that obtained
with the original model Hamiltonian.  The stationary current obtained with 4 SPFs is 20\% larger, which
does not satisfy our convergence criterion but is still reasonable.   The single configuration (1 SPF) 
result is approximately four times as large as the true result.  For this bias voltage the artificial
correlation effect in the original model becomes quite large.  It requires 60 SPFs per electronic SP group
to obtain the converged result in Fig.~\ref{fig2}a for the original model.

For an even higher bias voltage of $V =1$V, it becomes difficult to converge the steady-state current for the
original model due to the even increased artificial electronic correlation.  On the other hand, this also
makes the vibrational contribution (relatively) less important.  Figure~\ref{fig2}b shows
that within the scattering state
representation of the  Hamiltonian, the steady state current is easily
converged,  although the transient behavior is 
different with different number of SPFs.  In this case, the single
configuration result is very accurate, thus the steady-state current is very well
approximated by the purely electronic model. This is due to the rather
large voltage. It is known that  in the limit
of large voltages (within the wide-band-approximation), the steady-state current becomes
independent of the vibronic coupling.\cite{Gurvitz96} For the voltage
considered here, the polaron shifted energy level ($E_d
- \lambda \approx 0.25$ eV) is already well within the
bias window given by the chemical potential of the two electrodes.

\subsection{Vibrationally coupled electron transport for $E_d=E_f$}

Here we consider a model where the energy of the bridge state is located at the Fermi energy of the leads.  
This parameter regime is particularly interesting, because already for small bias voltage the transport 
mechanism corresponds to resonant tunneling and involves mixed electron/hole
transport. 

\begin{figure}[!ht]
\begin{flushleft}
(a)
\end{flushleft}
\includegraphics[clip,width=0.45\textwidth]{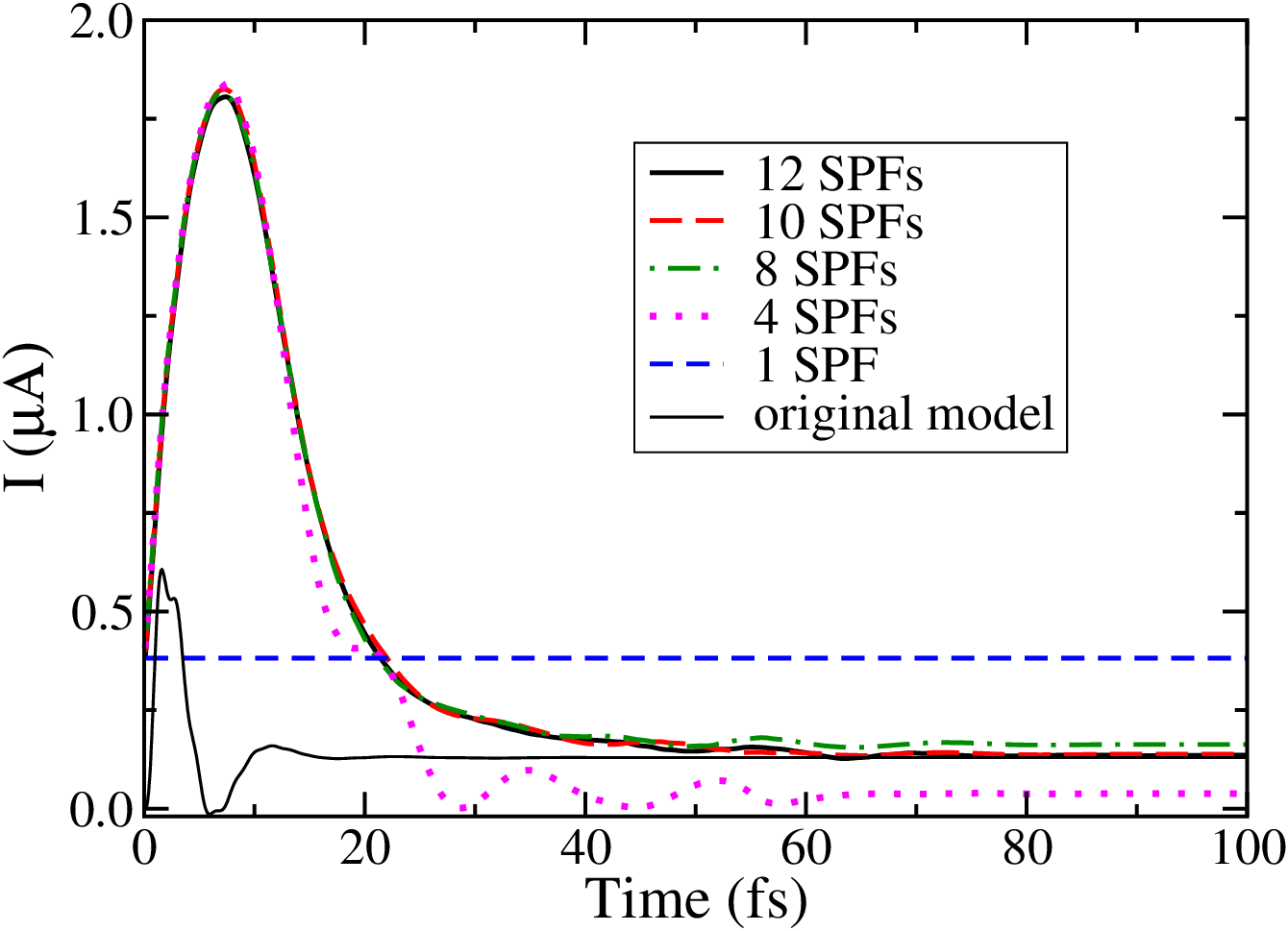}

\begin{flushleft}
(b)
\end{flushleft}
\includegraphics[clip,width=0.45\textwidth]{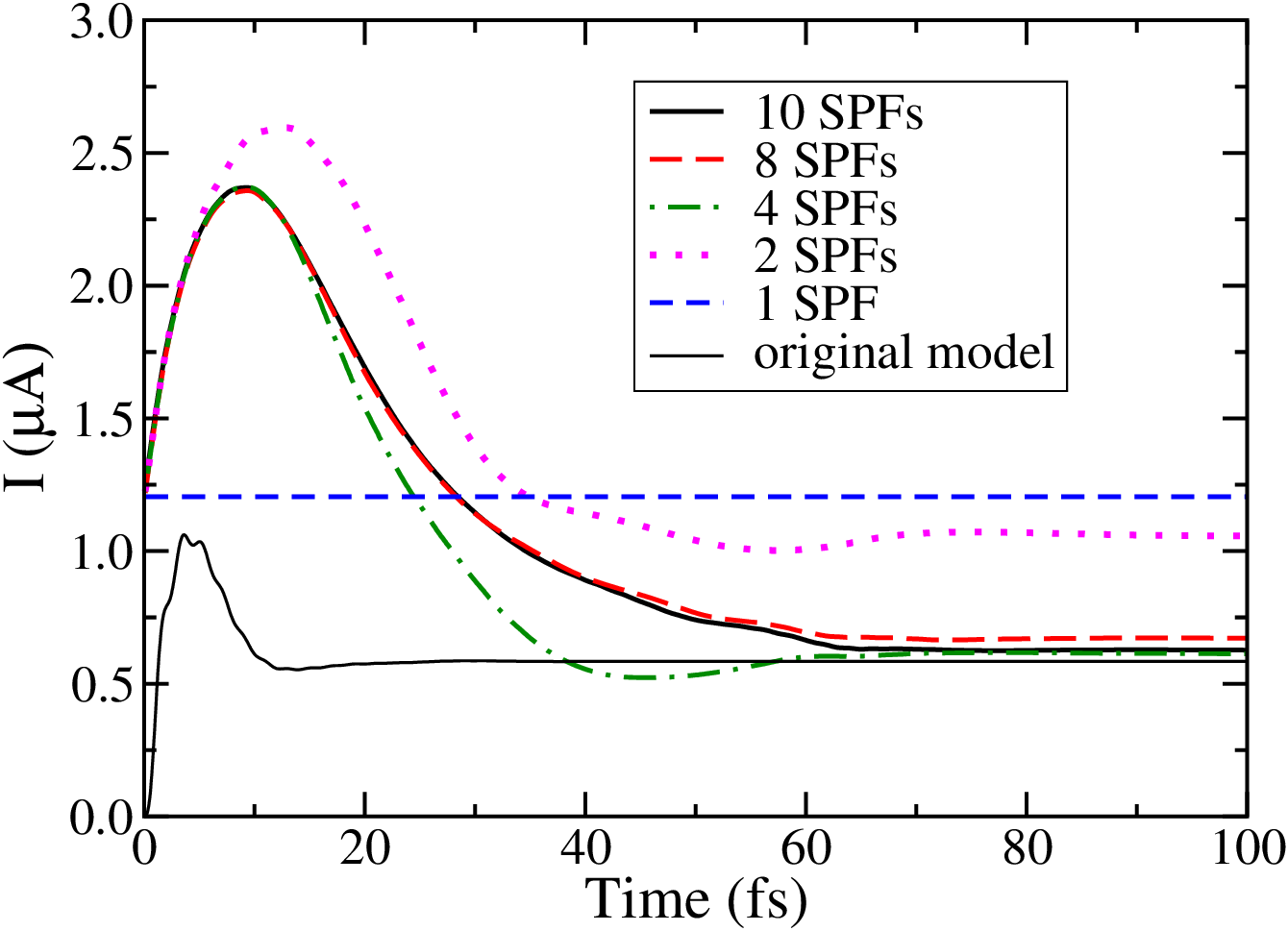}

\caption{Time-dependent current $I(t)$ versus the number of SPFs 
for the source-drain voltage $V = 0.1$V.  Six SPFs are used for each nuclear SP group.
The model parameters are: $\alpha_e = 0.2$eV, 
$\beta_e = 1$eV, and $E_d - E_f = 0$ for the electronic part; 
$\omega_c = 500{\rm cm}^{-1}$ for the vibrational bath and (a) $\lambda = 2000{\rm cm}^{-1}$,
(b) $\lambda = 1000{\rm cm}^{-1}$. 
}
\label{fig3}
\end{figure}

In this parameter regime and for a relatively large reorganization energy 
(e.g., $\lambda = 2000~{\rm cm}^{-1}$), it is found that when transforming to the effective
Hamiltonian in the scattering state representation, the naive procedure outlined in Sec.~\ref{naive} 
sometimes gives negative steady-state currents, which is unphysical.  This numerical problem persists 
even with a relatively large number of scattering channels and a large
configuration space.  Employing the modification
described in Sec.~\ref{specific},  on the other hand, 
gives excellent performance.  As shown in Figure~\ref{fig3} for a typical range of bath reorganization
energies, 1000-2000~cm$^{-1}$, results obtained with 10 SPFs per electronic SP group are in good agreement 
with those obtained with the original model Hamiltonian, which requires 40 SPFs per electronic SP group.  

\begin{figure}[!ht]

\includegraphics[clip,width=0.45\textwidth]{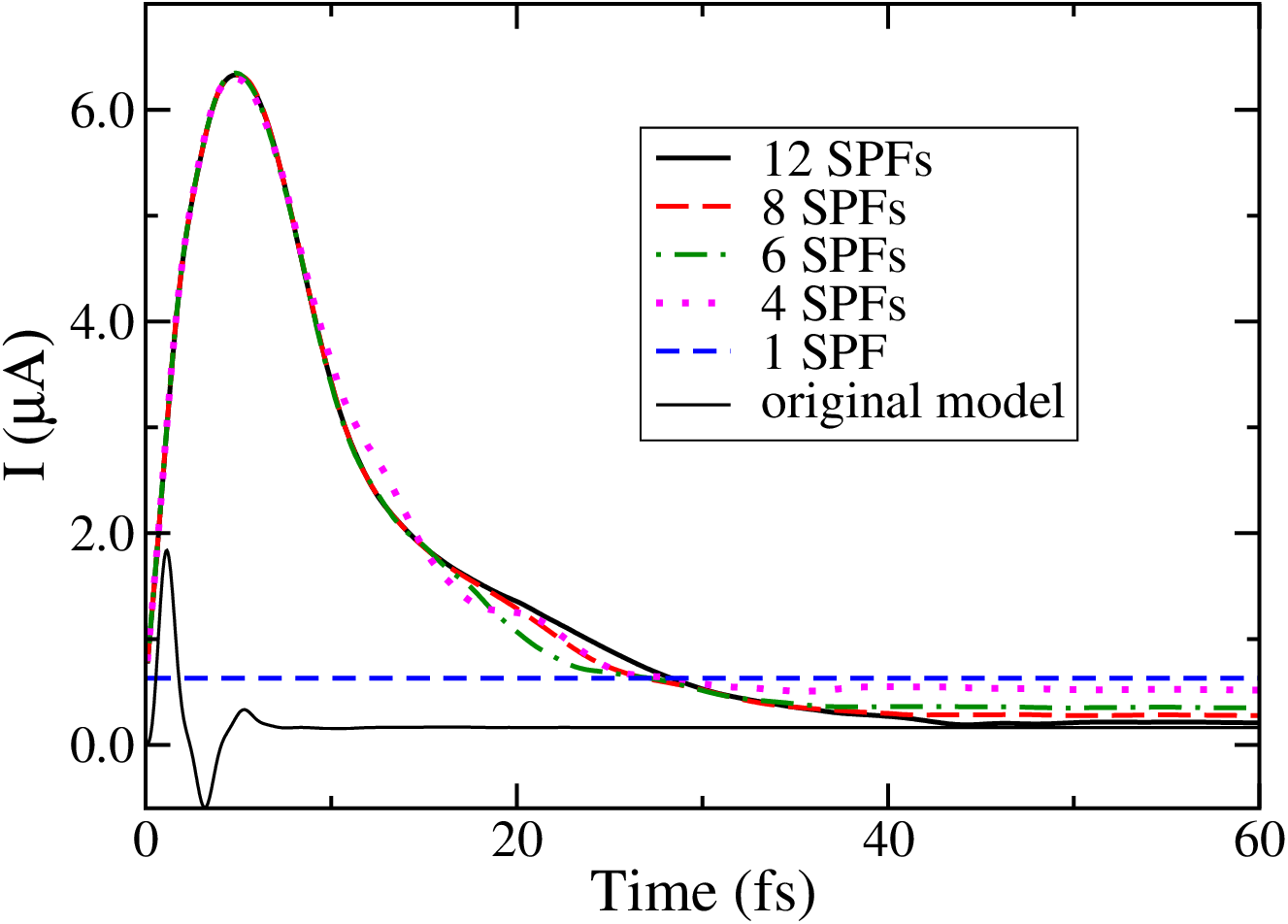}

\caption{Time-dependent current $I(t)$ versus the number of SPFs 
for the source-drain voltage $V = 0.5$V. Six SPFs are used for each nuclear SP group.
The model parameters are: $\alpha_e = 0.2$eV, 
$\beta_e = 1$eV, and $E_d - E_f = 0$ for the electronic part; 
$\omega_c = 500{\rm cm}^{-1}$ for the vibrational bath and $\lambda = 4000{\rm cm}^{-1}$.}
\label{fig4}
\end{figure}

The results depicted in Figs.~\ref{fig3}, \ref{fig4} for different values of
voltage and electron-vibrational coupling strength, all show the same
qualitative behavior: a strong increase
of the current for short time, which is followed by a decrease to a very small
steady-state current.  As was discussed in detail in in Ref.\ \onlinecite{Wang11},
this suppression of the steady state current is a manifestation of the phonon
blockade effect, which  is particularly
pronounced for larger electronic-vibrational coupling. 

The phonon blockade of the current is usually, qualitatively rationalized by considering the
energy level of the bridge state. For any finite bias voltage, the bare energy of the
bridge state ($E_d - E_f = 0$) is located between  the chemical potential of the leads and thus, 
within a purely electronic model, current can flow. The coupling to the vibrations results in a 
polaron shift of the energy of the bridge state. For electronic-vibrational coupling strengths  
$\lambda > |V|/2$ the polaron-shifted energy of the bridge state is below the chemical potentials 
of both leads and thus the current is blocked.

The comparison between the converged results and those obtained employing only
a single SPF for the electronic degrees of freedom (which agree with the
results for the purely electronic model), demonstrate that this simple picture
based on the static energy shift is at best qualitatively correct. It is seen
from Fig.~\ref{fig3} that  the 1 SPF (single configuration) approach, i.e., the Hartree-Fock 
limit in the scattering state representation, significantly overestimates the
steady state current.   The discrepancy becomes larger 
for stronger electron-nuclear couplings.

It is noted that the results obtained with 1 SPF, which neglect true correlation
between electronic and vibrational degrees of freedom, are very similar
to the results obtained with a NEGF
method\cite{Galperin06,Haertle08} (see, e.g., the data shown in Fig.~9 of
Ref.\ \onlinecite{Wang11}). 
In this method, a factorization of the
electronic and vibrational Green's function is employed. As a consequence,
correlations between electronic and vibrational degrees of freedom are only
treated on a very approximate level.  This results, in this case, in a significant
overestimation of the current, similar to what would be obtained with a very
small (unconverged) number of configurations in the ML-MCTDH-SQR method.

Similar to the examples discussed in Sec.\ \ref{large_Ed}, the single configuration limit becomes a better 
approximation for a higher bias voltage when the vibronic coupling is fixed.  For the parameters 
displayed in Fig.~\ref{fig3}, the single configuration result becomes very good for, e.g., a bias
voltage of 1 V (data not shown), similar to that illustrated in
Fig.~\ref{fig2}b.  Again, in this regime, the NEGF 
approach or the Landauer formula with the shifted bridge state energy becomes accurate.

It should be emphasized, though, that this also depends on the relative strength
between the vibronic coupling  and the bias 
voltage. Only for voltages $|V|\gg |2(E_d \pm \lambda - E_f)|$, can it  be
expected that vibronic effects, and thus also vibrationally induced
correlation, are negligible  in the steady state current.\cite{Gurvitz96}  As an example, Figure~\ref{fig4}a shows the simulation for a higher voltage of 0.5 V but also a larger
reorganization energy 4000~cm$^{-1}$.  In this case 12 SPFs per electronic SP group are required to
obtain the result in the scattering state representation, where the single configuration result is 
approximately a factor of three too large.   Convergence in the original model Hamiltonian
also becomes more expensive, which requires 64 SPFs per electronic SP group.

\subsection{Vibrationally coupled electron transport with $E_d < E_f$}

\begin{figure}[!ht]

\includegraphics[clip,width=0.45\textwidth]{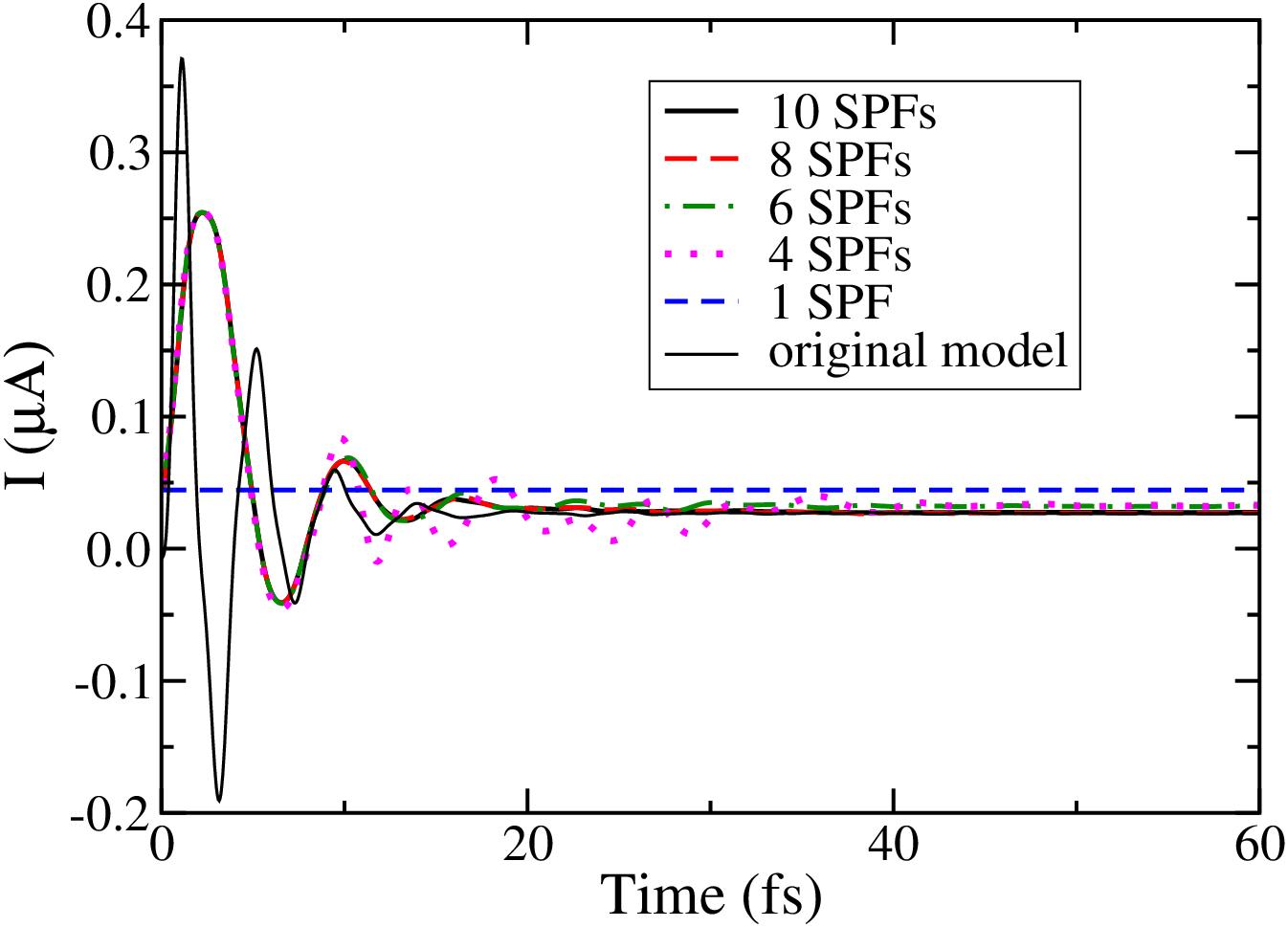}

\caption{Time-dependent current $I(t)$ versus the number of SPFs 
for the source-drain voltage $V = 0.1$V. Six SPFs are used for each nuclear SP group.
The model parameters are: $\alpha_e = 0.2$eV, 
$\beta_e = 1$eV, and $E_d - E_f = -0.5$eV for the electronic part; 
 $\lambda = 2000{\rm cm}^{-1}$ and $\omega_c = 500{\rm cm}^{-1}$ for the vibrational bath.}
\label{fig5}
\end{figure}

We finally consider in Fig.~\ref{fig5} an example with a bridge state located at $E_d - E_f=
-0.5$ eV., i.e. well  below the Fermi energy.  
All other parameters are the same as in Fig.~\ref{fig1}. For the voltage
considered, this case  corresponds to non-resonant transport.
Similar to the case $E_d - E_f = 0$,  when transforming to the effective Hamiltonian in the scattering 
state representation, the naive procedure outlined in Sec.~\ref{naive} gives a
negative steady-state current.  On the other hand, Figure~\ref{fig5} shows the result
obtained with employing  the modification described in Sec.~\ref{specific},
which gives excellent  agreement with the result obtained with
the original model Hamiltonian.  The former requires 8 SPFs to converge, whereas the latter require
40 SPFs.  The performance with other voltage/vibronic coupling is similar to the previous figures.

\section{Concluding Remarks}\label{conclusions}

In this paper, we have shown that the scattering state representation of the
underlying electronic Hamiltonian represents a promising approach for studying
vibrationally coupled  electron transport through molecular junctions. 
Within the ML-MCTDH-SQR theory\cite{wan09:024114,Wang11} much less configurations are required to achieve 
convergence than when using the original representation of the Hamiltonian.  This is because
the transformation to the
scattering state representation effectively diagonalizes the underlying
electronic Hamiltonian and, therefore,  much artificial correlation present in the original model is removed.
We have also shown that the definition of the scattering states is not unique.
This can be used to optimize the simulation. Our numerical studies suggest
that the form of the Hamiltonian outlined in 
Sec.~\ref{specific} performs best in all the physical regimes investigated.

The results of our studies also give physical insight into the importance of
correlation effects in vibrationally coupled nonequilibrium transport. While
for small voltages and/or transient effects,
electronic-vibrational correlation is often of utmost importance, for higher
voltages correlation effects are less pronounced in the steady state current.  
This finding is also of relevance for the validity of NEGF calculations for
vibrationally coupled electron transport, in particular, methods that employ
an effective factorization between electronic and vibrational degrees of freedom.\cite{Galperin06,Haertle08}

To conclude this paper, we discuss the computational costs of the simulations
in the two different representations.
As was shown above, much  less SPFs are needed when
using the scattering state representation as compared to the original model.
However, the transformation  used to generate the number 
operator $d^+d$ and the current operator $\hat{I}_\zeta$, Eq.~(\ref{backward}), scales quadratically
versus the number of the electronic states.  This is in contrast to the original model where the number
of electronic operators scales linearly versus the number of electronic states.
As a result, there are many more operators to evaluate in the scattering state Hamiltonian, which makes
the simulation more expensive.  For the examples considered in this paper, using 10 SPFs per electronic
SP group in the scattering model has a similar computational cost to using 40 SPFs in the original model.
Therefore, when studying vibrationally coupled electron transport at a lower bias voltage, the original 
model Hamiltonian may be more efficient.

However, things are quite different when the bias voltage becomes higher, where significant artificial
correlation exists in the original model, which makes convergence difficult or
even impossible.  The scattering state representation, on the other hand, requires
typically a smaller configuration space 
to converge as demonstrated by the numerical examples
in this paper.  This makes the scattering state representation a valuable tool for multiconfigurational
study of current-voltage characteristics where the bias voltage spans a larger range.  Such work is currently
in progress.

Our current implementation of the scattering state representation has not yet exploited its full potential.
It is possible that a different energy shift from that of the reorganization energy is more efficient for
solving the problem in some physical regimes.  This may be achieved by designing a self-consistent 
procedure in the convergence tests.  Furthermore, a different initial condition for the
vibrational bath or even a correlated electronic-vibrational initial density matrix may be helpful
in reducing the transient peaks of the current, and thus making the calculation of the steady-state 
current more stable numerically.  Work on these issues is in progress.

\section*{Acknowledgments}
This work has been supported by the National Science Foundation
CHE-1012479 (HW), the 
German-Israeli Foundation for Scientific Development (GIF) (MT), and the 
Deutsche Forschungsgemeinschaft (DFG) through SFB 953 and the Cluster of Excellence ’Engineering of Advanced
Materials’ (MT), and used resources
of the National Energy Research Scientific Computing Center, which is supported by the
Office of Science of the U.S.  Department of Energy under Contract
No. DE-AC02-05CH11231.

\pagebreak

%
%


\clearpage

\clearpage

\clearpage

\clearpage
~
\vspace{3cm}

\clearpage
~
\vspace{3cm}

\end{document}